\documentclass[11pt]{article}

\usepackage[letterpaper,margin=1in,headheight=16pt]{geometry}
\usepackage[T1]{fontenc}
\usepackage{lmodern}
\usepackage{microtype} 
\usepackage{graphicx}
\usepackage{booktabs}
\usepackage{enumitem}
\usepackage{titlesec}
\usepackage{fancyhdr}
\usepackage{amsmath}
\usepackage{amssymb}
\usepackage[square,numbers]{natbib}

\newcommand{\apjl}{Astrophysical Journal Letters}
\newcommand{\apjs}{Astrophysical Journal Supplements}

\newcommand{\aap}{Astronomy \& Astrophysics}

\newcommand{\ergs}{\ifmmode \textrm{erg\ s}^{-1} \else erg s$^{-1}$\fi}
\newcommand{\Msun}{\ifmmode \textrm{M}_{\odot} \else $M_{\odot}$\fi}
\newcommand{\Lsun}{\ifmmode \textrm{L}_{\odot} \else $L_{\odot}$\fi}

\usepackage{xcolor}
\definecolor{genesisgold}{HTML}{B5771A}
\definecolor{genesisink}{HTML}{10131F}
\definecolor{genesisgray}{HTML}{6B7386}

\usepackage[colorlinks=true,linkcolor=genesisgold,citecolor=genesisgold,urlcolor=genesisgold]{hyperref}

\titleformat{\section}{\normalfont\large\bfseries\color{genesisink}}{}{0pt}{}
\titlespacing*{\section}{0pt}{1.3em}{0.45em}

\fancypagestyle{genesisfirst}{%
  \fancyhf{}%
  \fancyhead[L]{}%
  \fancyhead[R]{\small\textcolor{genesisgray}{MCT Concept White Paper}}%
  \fancyfoot[C]{\small\thepage}%
}

\begin{document}
\thispagestyle{genesisfirst}

{\LARGE\bfseries\color{genesisink} What observations are needed to advance our theoretical understanding of galaxy formation at high redshift?\par}
\vspace{0.8em}

{\large H.~Katz, R.~S.~Somerville, R.~Schneider, B.~E.~Robertson, F.~Pacucci, and 96 co-signers\par}
\vspace{0.3em}
{\small\itshape\textcolor{genesisgray}{The full list of authors, co-signers and affiliations is given on page~\pageref{sec:authors}.}\par}
\vspace{0.4em}

{\small Corresponding author: H.~Katz (\href{mailto:harleykatz@uchicago.edu}{harleykatz@uchicago.edu})\par}

\vspace{1.0em}

\noindent\textbf{\color{genesisink}Abstract.}\enspace
Although significant progress has been made over recent decades, a comprehensive and predictive framework for galaxy formation remains elusive. Starting from cosmological initial conditions, current models cannot \emph{a priori} accurately predict key galaxy properties such as their structure, size, star formation histories, chemical abundance patterns, or supermassive black hole mass. This challenge stems from the vast range of spatial scales involved, as numerical simulations must simultaneously account for sub-AU processes nearby stars and black holes in addition to Mpc-scale cosmological gas inflows/outflows and environmental effects. Consequently, simulations must rely on subgrid models and approximations. As a result, theoretical progress in galaxy formation remains fundamentally driven by observational data and the iterative testing of model predictions against empirical results. Here we highlight five key open problems in theoretical galaxy formation that require dedicated observations of the high-redshift Universe, beyond what can currently be achieved through GO programs. 

\section{Background, Scientific Justification, \& Requests from the Theory Community}

Within the first few cycles of JWST observations, our understanding of early chemical enrichment, star formation modes, ISM conditions, and black holes has been upended by a surprising wealth of observations that conflict with many of the expectations from pre-JWST models. While this can be perceived negatively as a failure of these models, rather, as a theoretical community we view these disagreements as an important opportunity to learn about our cosmos and improve our understanding of the key physics driving early galaxy formation. As we recognize that we are not yet in a position to “solve” galaxy formation from first principles, we are reliant on the observational community to not only provide benchmarks to test our models\footnote{``Models'' here is meant in a very general sense, including cosmological simulations, semi-analytic models, semi-empirical models, etc.} against, but also to fill the gaps in our understanding of the key physical processes we may be missing. For this reason, as a community, we are extremely supportive of dedicated observations of early galaxy formation, probing current and emerging frontiers, that both settle debates as well as “break” our models in new and exciting ways. More specifically, we argue that further progress requires
\begin{enumerate}
\item Statistically robust samples of galaxies with \emph{well-understood selection functions}, \emph{uniform observation strategies}, and \emph{uniform coverage from the UV to near-IR}.

\item Larger samples of \emph{high-resolution spectroscopy} from which physical properties (chemical abundances, star formation histories, ISM conditions, etc.) can be robustly estimated.

\item \emph{Significantly deeper imaging and spectroscopy} that advance the low-mass and high-redshift frontier at Cosmic Dawn.

\end{enumerate}

Obtaining these samples will require resource-intensive observational campaigns that are significantly more ambitious than what is currently supported through GO programs. As JWST is the only facility that can achieve our goals, below we outline five key areas where empirical guidance would have the greatest impact on our work.

\begin{figure}
    \centering
    \includegraphics[width=0.35\linewidth]{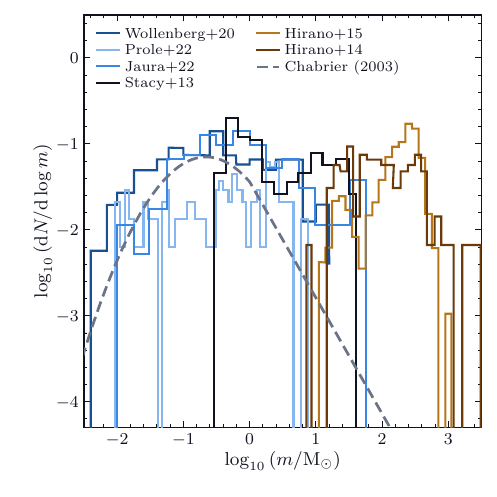}
    \includegraphics[width=0.35\linewidth]{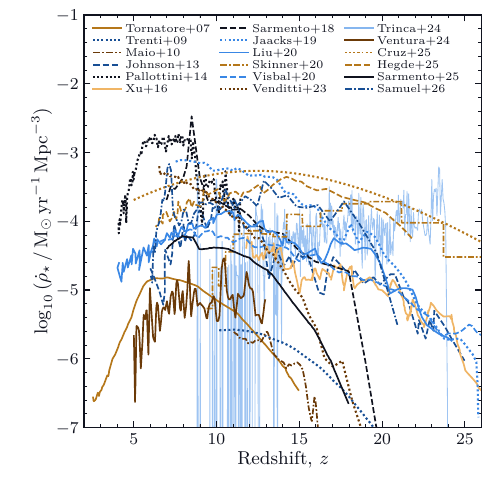}
    \vspace{-5mm}
    \caption{Predictions for the Pop~III IMF ({\it left}) and Pop~III star formation rate density as a function of redshift ({\it right}) for various semi-analytic models and simulations \cite{Wollenberg+20,Prole+22,Jaura+22,StacyBromm13,Hirano+15,Hirano+14,Chabrier03,Tornatore+07,TrentiStiavelli09,Maio+10,Johnson+13,Pallottini+14,Xu+16a,Sarmento+18,Jaacks+19,SkinnerWise20,Liu2020,Visbal+20,Venditti+23,Trinca+24,Ventura+24,Cruz+25,HegdeFurlanetto25,SarmentoScannapieco25,Samuel+26}. These figures were adapted from \cite{Venditti2026}.}
    \label{fig:p3}
\end{figure}

\subsection{1. Uncovering the Properties of Pop III and Low-Metallicity Massive Stars}
While very little is known about massive stars at low metallicity, primarily due to a lack of low-redshift environments where such stars can be probed in detail, the physical properties of Pop~III stars remain almost completely unconstrained. Empirical constraints on both of these elusive stellar populations is absolutely paramount for our galaxy formation models. Enrichment from Pop~III stars drives the transition from metal-free to metal-enriched gas \citep[e.g.][]{Omukai2005,Schneider2012} and may be imprinted on a population of ultra-faint dwarf galaxies that survive today \cite{Frebel2005,Chiti2026}. Pop II star formation is believed to be much more efficient due to metal-line and dust cooling and so understanding when and how quickly this transition occurred is absolutely key for deciphering why there exists a population of massive, mature galaxies in the early Universe. Currently, simulations differ significantly on the Pop~III star formation rate density as a function of redshift, when Pop~III star formation ends, and what the Pop~III IMF is (See Figure~\ref{fig:p3}), and the role of Pop~III stars in reionization. While quantitative values disagree, almost all simulations predict that the Pop III IMF differs from that of Pop II stars and most simulations now predict ``late'' Pop~III star formation at redshift $z<10$. Only empirical data can settle these quantitative debates.

More generally, massive stars dominate the energetic feedback budget of galaxies via their radiation (e.g. direct radiation pressure, photoheating, Ly$\alpha$ radiation pressure) and mechanical feedback (e.g. supernova, stellar winds), and drive how galaxies regulate their stellar masses and contribute to reionization \cite{Leitherer1999,Agertz2013}. Their spectral properties, how they evolve, and their chemical yields remain highly uncertain inputs into current models. 

First, we recommend significant time be invested in achieving the first definitive detection of Pop~III stars. Almost all numerical simulations predict that Pop~III galaxies are faint and short-lived, and thus this would require significant exposure times, most likely over lensing fields. Indeed, there are tantalizing hints from existing observations of objects with features consistent with Pop~III predictions \cite{Maiolino2026}. Going deeper should then reveal even more Pop~III-like objects. A single robust Pop~III star detection would not only have paradigm-shifting consequences for our understanding of Cosmic Dawn, it would fulfill one of the primary science cases of the JWST mission. Beyond individual detections, statistical constraints on the abundance, environments, and the redshift evolution of Pop~III star-forming systems would be crucial for distinguishing between competing models of primordial star formation and the transition to metal-enriched stellar populations. Second, we recommend that significant additional time be invested in uncovering the properties of massive Pop II stars at high redshift. Deep rest-frame UV spectra have already revealed that their properties and chemical enrichment patterns differ significantly from the current inputs of stellar population synthesis models \cite{Chen2026,Nakane2026}. Further observations are thus key to refine our models.

\subsection{2. Reionization Quenching at the Faint End of the UV Luminosity Function}
\begin{figure}
    \centering
    \includegraphics[width=0.35\linewidth]{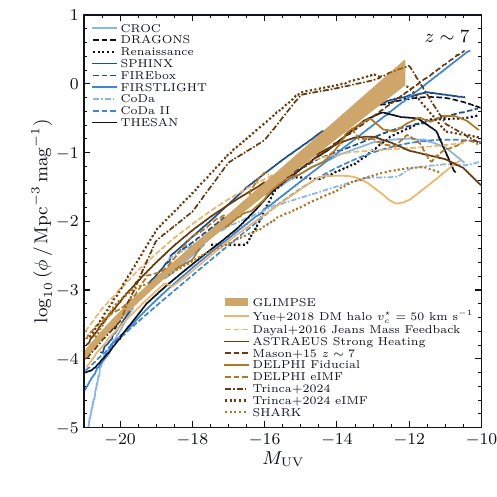}
    \includegraphics[width=0.35\linewidth]{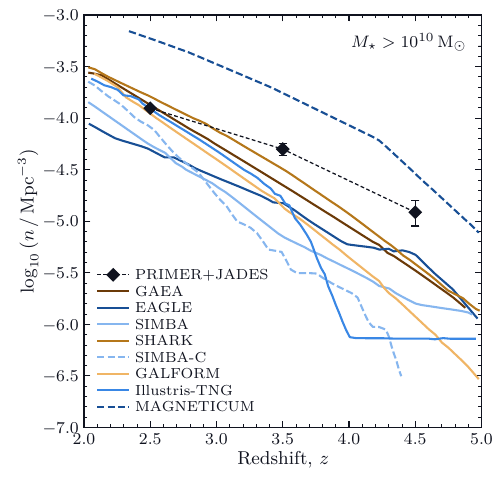}
    \vspace{-5mm}
    \caption{({\it Left}) Comparison of 1500~\AA\ UVLFs from various models \cite{Gnedin2016,Liu2016,Oshea2015,Rosdahl2018,Ocvirk2016,Ocvirk2020,Dawoodbhoy2023,Kannan2022,Ceverino2017,Feldmann2023,Hutter2021,Mauerhofer2025,Trinca2024,Mason2015,Dayal2015,Finlator2017,Yue2018} at $z\sim7$ to empirical constraints from GLIMPSE \cite{Atek2026}, which provides the deepest current constraints. While the models converge at the bright end, significant disagreement remains at the faint end. ({\it Right}) Comparison of quenched number density predictions from theoretical models \cite{Lagos2024,Schaye:2015,Dave2019,DeLucia2024,Lacey:2016,Pillepich2018,Dolag2025,Hough2023} to observational constraints from PRIMER and JADES \cite{Stevenson2026}. These figures were adapted from \cite{Atek2026} and \cite{Stevenson2026}, respectively.}
    \label{fig:UVLF}
\end{figure}

A robust prediction of cosmological simulations is a turnover in the faint end of the UV luminosity function (UVLF), whose location and shape depend on the interplay between radiative feedback from reionization, stellar feedback, and the efficiency of star formation in low-mass halos \citep[e.g.][]{Sobacchi2013,Katz2020}. While this prediction is nearly ubiquitous, significant disagreement remains regarding at what halo mass the turnover should occur, how inhomogeneous radiation impacts the turnover, and whether for radiative feedback strangulation versus photoevaporation leads to a gradual versus an instantaneous formation of the turnover (see Figure~\ref{fig:UVLF}). Empirically constraining the turnover, before, during, and after reionization will help disentangle how feedback processes operate in low-mass galaxies and would provide an important lens through which to understand local dwarf galaxies. More generally, almost all post-JWST models for galaxy formation have now been tuned to match the surprising over-abundance of UV bright galaxies, but these models still differ significantly in their behavior at the faint end, below the current detection limits ($M_{\rm UV}\lesssim -13$) of JWST.

For these reasons, we recommend a dedicated set of observations, specifically designed to surpass the current faint frontier in the UVLF set by the Hubble and JWST \cite{Atek2018,Bouwens2022b,Atek2026}, and probe as deep as possible to constrain the faint-end slope and the turnover, before, during, and after reionization.  

\subsection{3. Galaxy Scaling Relations as Crucial Tests of Galaxy Formation Physics}
\looseness=-1 
While numerical simulations provide self-consistent frameworks for galaxy evolution, direct comparisons of model results with any  individual galaxy is fundamentally limited by stochasticity, cosmic variance, and unresolved physics. Rigorous tests of our theoretical models must therefore be conducted at the population level. Galaxy scaling relations such as the mass–metallicity relation, the fundamental metallicity relation, stellar mass-halo mass relation, and the star-forming main sequence are now empirically constrained at high redshift \citep{Curti2024,Laseter2026,Clarke2026}. Because these relations emerge from the complex interplay of gas accretion, star formation, and feedback, they serve as decisive tests of theoretical model predictions. However, the primary issue with current comparisons of theoretical and observed galaxy scaling relations is that we are comparing apples to oranges, kiwis, and grapes.

Simulations have made enormous strides in their ability to forward-model observations and infer galaxy properties in the same way an observer would from JWST images or spectra \citep[e.g.][]{Choustikov2026}. There are now numerous public codes (e.g., SKIRT \cite{Camps2020}, COLT \cite{McClymont2025}, RASCAS \cite{MD2020}, SYNTHESIZER \cite{Lovell2025}) that allow simulators to make these comparisons. However, particularly in the context of MSA spectroscopy, very few observational programs have a well-defined selection function that can be replicated by simulations. This problem is further exacerbated by meta-analyses, where multiple, ill-defined selection functions are combined into a single, published galaxy scaling relation. Ill-defined selection functions and spectroscopic inhomogeneities (e.g., varying spectral resolutions and integration times) currently represent the primary impediment to utilizing scaling relations as constraints on theory. Well-characterized selection functions, coupled with a standardized framework to forward-model observables and infer galaxy and ISM physical properties, is therefore paramount if JWST scaling relations are to serve as rigorous benchmarks for theoretical models.

We therefore recommend a dedicated, carefully planned, and coordinated spectroscopic survey at $z>4$, particularly designed to constrain the mass-metallicity relation, fundamental metallicity relation, and star formation main-sequence, with a well-defined selection function and a homogeneous observing mode that can be robustly replicated by mock observations of simulated galaxies.  

\subsection{4. Constraining Modes of Star Formation and Feedback Physics at Cosmic Dawn}
JWST has now discovered hundreds of photometric candidates and $>40$ spectroscopically confirmed $z \gtrsim 10$ galaxies \citep[e.g.][]{RobertsBorsani2026}, which is significantly in excess of what was predicted by most pre-launch models and simulations \citep[see][for a summary]{Ellis_saasfee2025,somerville_saasfee2026}. The main explanations for this discrepancy are 1) galaxies were brighter at early times, either because of lower dust attenuation \citep{Ferrara2023,Narayanan:2025b,Sommovigo2026}, differences in stellar populations (e.g. a ``top heavy'' stellar initial mass function richer in massive stars; \citep{Trinca2024}), bursty star formation \citep{Sun2023}, or a contribution from accreting black holes \citep{Trinca2024} or, 2) star formation was more efficient in the early Universe, perhaps due to weaker stellar feedback and shorter dynamical times \citep{Dekel2023,Somerville:2025}. An additional complication is the possible contribution of accreting black holes to the observed UV emission, which may bias the inferred stellar masses and star formation efficiencies if not properly accounted for. Disentangling stellar and AGN contributions is therefore essential for interpreting the bright end of the UVLF and testing competing galaxy formation scenarios. In order to empirically constrain the importance of these different physical effects, we need deep high-resolution spectroscopy, which can be used to infer star formation histories and burstiness as well as AGN signatures or IMF variations, and deeper imaging to measure galaxy shapes/sizes and push UVLF measurements to even higher redshifts, where existing models make very different predictions \citep{Weibel2025}. Probing large-scale clustering via pure parallel imaging is another strategy to disentangle these proposed mechanisms \citep{Weibel2026}. 

\subsection{5. Which Physical Mechanisms Drive Quenching in Massive Galaxies?}
In addition to struggling to reproduce the bright end of the UVLF at high redshift, most pre-JWST theoretical models cannot produce the population of massive quiescent galaxies at lower redshifts (e.g. $3\lesssim z\lesssim5$, see right panel of Figure~\ref{fig:UVLF}). In fact, some estimates of the quiescent galaxy population at $z\sim7$ are $100\times$ higher than model predictions \cite{Weibel:2025} while the mass and star formation histories of lower redshift quiescent galaxies have been proposed as a challenge to $\Lambda$CDM \cite{Glazebrook2024}. Empirical constraints on the stellar masses, star formation histories, and chemical abundance patterns are crucial for elucidating quenching mechanisms, and informing how feedback and gas accretion operate in some of the most extreme environments in our Universe. While AGN feedback is likely the dominant mechanism for quenching massive galaxies, the importance of other physics such as virial shocks and ram-pressure stripping is unknown. Within the context of AGN feedback, the exact role of winds driven by radiatively efficient accretion (``quasar mode'') vs. relativistic jets powered by radiatively inefficient accretion  (``radio mode'') also remain debated. This has lead to widely disparate predictions for quenched galaxy number densities between numerical simulations and semi-analytic models \cite[see right panel of Figure~\ref{fig:UVLF},][]{Lagos2025,Stevenson2026,Baker2025}. 

\looseness=-1 
A key issue is that large degeneracies exist in star formation histories that limit our ability to constrain the underlying physics \cite{degraaff2025}. Such degeneracies can be broken with deep, high-resolution 2D spectroscopy that can simultaneously constrain dynamical mass and stellar abundance patterns and gradients, both of which inform the star formation history and thus the quenching timescale. As these objects are rare, systematic surveys with deep (e.g. IFU) spectroscopy, perhaps based on targets selected from Euclid and Roman will be required to reconcile the simulations with observations.

\section{Summary}
The five science cases above represent the most pressing problems faced by the extragalactic theory community\footnote{We have explicitly focused here on galaxies. A companion discussion on the open theoretical questions related to black holes is presented in \cite{genesistheory2} and this work falls under the Genesis Initiative described in \cite{genesis}.}. These are directly related to the key scientific drivers of JWST: ``The End of the Dark Ages: First Light and Reionization'' and ``The Assembly of Galaxies'' \cite{Gardner2006}. Some of these open questions (e.g. the nature of Pop~III stars) were well established prior to the launch of JWST, while new problems (e.g. the abundance of bright, high-redshift galaxies) emerged as a result of the fantastic early achievements of the telescope. Nevertheless, solving these will require significant investments in observations that are not achievable with current GO programs. Addressing the breadth of these challenges may require distinct, tailored observational strategies including:
\begin{enumerate}
    \item Deep spectroscopy in many lensing fields to measure the UVLF at the faint end at high redshift and detect Pop~III stars.
    \item Deep, wide-field, homogeneous imaging and spectroscopy to constrain galaxy scaling relations across a wide range in stellar mass and the UVLF at extremely high-redshifts.
    \item Deep, high-resolution (and potentially IFU) spectroscopy for a large sample of massive galaxies to constrain the physics of quenching.
\end{enumerate}
Addressing any of these science cases through a dedicated MCT program will have an immense, enduring impact on the high-redshift theory community.

\clearpage
\section{Authors -- \textcolor{genesisgold}{101 Authors and Co-Signers from the Theory Community}}\label{sec:authors}
\noindent
H.~Katz\textsuperscript{1},
R.~S.~Somerville\textsuperscript{2},
R.~Schneider\textsuperscript{3},
B.~E.~Robertson\textsuperscript{4},
F.~Pacucci\textsuperscript{5},
Abdurro'uf\textsuperscript{6},
K.~Aditya\textsuperscript{7},
J.~M.~Afonso\textsuperscript{8,9},
M.~Aravena\textsuperscript{7},
G.~Barro\textsuperscript{10},
A.~Basu\textsuperscript{11},
R.~Begley\textsuperscript{12},
S.~Bhattacharya\textsuperscript{13},
S.~Bose\textsuperscript{14},
M.~Brada\v{c}\textsuperscript{15,16},
G.~L.~Bryan\textsuperscript{17},
C.~Cain\textsuperscript{18,19,20},
A.~J.~Cameron\textsuperscript{21,22},
S.~Carniani\textsuperscript{23},
C.~M.~Cheng\textsuperscript{24},
N.~Choustikov\textsuperscript{18,19},
E.~Curtis-Lake\textsuperscript{13},
A.~D'Angelo\textsuperscript{25,26,3,27},
R.~Dav\'e\textsuperscript{28},
R.~L.~Davies\textsuperscript{29},
F.~van Donkelaar\textsuperscript{19,18},
C.~T.~Donnan\textsuperscript{30},
N.~E.~Drakos\textsuperscript{31},
Q.~Duan\textsuperscript{18},
Y.~Dubois\textsuperscript{32},
A.~L.~Faisst\textsuperscript{33},
M.~Farcy\textsuperscript{34},
S.~L.~Finkelstein\textsuperscript{35,36},
A.~Fontana\textsuperscript{25},
S.~Gillman\textsuperscript{37},
S.~C.~O.~Glover\textsuperscript{38},
L.~Graziani\textsuperscript{3},
A.~R.~Gupta\textsuperscript{35,36},
T.~A.~Gutcke\textsuperscript{39},
M.~G.~Haehnelt\textsuperscript{18,19},
N.~P.~Hathi\textsuperscript{40},
K.~E.~Heintz\textsuperscript{37},
M.~Hirschmann\textsuperscript{34},
M.~Huertas-Company\textsuperscript{41,42},
A.~Hutter\textsuperscript{43},
L.~R.~Ivey\textsuperscript{18},
K.~Iyer\textsuperscript{39},
J.~Jeon\textsuperscript{35,36},
T.~B.~Jeong\textsuperscript{36},
C.~K.~Jespersen\textsuperscript{44},
B.~L.~Jones\textsuperscript{45},
I.~Juod\v{z}balis\textsuperscript{46},
T.~Kimm\textsuperscript{47},
A.~M.~Koekemoer\textsuperscript{48},
D.~Langeroodi\textsuperscript{18},
A.~Lapi\textsuperscript{49},
M.~A.~Latif\textsuperscript{50},
Y.~Liu\textsuperscript{14},
C.~C.~Lovell\textsuperscript{18},
T.-Y.~Lu\textsuperscript{51},
F.~Mannucci\textsuperscript{52},
S.~H.~Menon\textsuperscript{2},
A.~Mesinger\textsuperscript{53},
K.~Nagamine\textsuperscript{54},
Y.~Nakazato\textsuperscript{55},
P.~Natarajan\textsuperscript{56,57,58},
O.~Nebrin\textsuperscript{59},
R.~Pascalau\textsuperscript{18},
B.~P\'erez-D\'iaz\textsuperscript{25},
P.~G.~P\'erez-Gonz\'alez\textsuperscript{60},
E.~Pizzati\textsuperscript{5},
A.~Polzin\textsuperscript{61,24},
E.~Regos\textsuperscript{62},
M.~Reina-Campos\textsuperscript{63},
M.~P.~Rey\textsuperscript{64},
P.~Rinaldi\textsuperscript{35,36},
D.~Roberts\textsuperscript{65},
F.~Rodr\'iguez Montero\textsuperscript{66},
W.~Roper\textsuperscript{67},
E.~Rusta\textsuperscript{37},
M.~Saggini\textsuperscript{3,26,27},
R.~L.~Sanders\textsuperscript{68},
F.~Shankar\textsuperscript{65},
A.~E.~Shapley\textsuperscript{69},
X.~Shen\textsuperscript{5},
T.~M.~Stanton\textsuperscript{28},
S.~D.~Stevenson\textsuperscript{28},
G.~Sun\textsuperscript{70},
J.~C.~Turner\textsuperscript{71},
A.~Venditti\textsuperscript{35,36},
A.~P.~Vijayan\textsuperscript{71},
T.~Wang\textsuperscript{72},
S.~Wilkins\textsuperscript{71},
J.~H.~Wise\textsuperscript{73},
J.~Witstok\textsuperscript{21,22},
Z.~Wu\textsuperscript{5},
L.~Y.~A.~Yung\textsuperscript{74},
T.~Zana\textsuperscript{3,25,27},
S.~Zhang\textsuperscript{36},
R.~J.~Zhao\textsuperscript{75,76},
M.-Y.~Zhuang\textsuperscript{77}

\vspace{0.6em}
{\small\itshape\emergencystretch=3em\textcolor{genesisgray}{%
\textsuperscript{1}Department of Astronomy \& Astrophysics, University of Chicago, 5640 S Ellis Avenue, Chicago, IL 60637, USA \quad
\textsuperscript{2}Center for Computational Astrophysics, Flatiron Institute, New York, NY 10010, USA \quad
\textsuperscript{3}Dipartimento di Fisica, Sapienza Universit\`a di Roma, Piazzale Aldo Moro 5, I-00185 Roma, Italy \quad
\textsuperscript{4}Department of Astronomy and Astrophysics, University of California, Santa Cruz, 1156 High Street, Santa Cruz, CA 95064, USA \quad
\textsuperscript{5}Center for Astrophysics $|$ Harvard \& Smithsonian, 60 Garden St., Cambridge, MA 02138, USA \quad
\textsuperscript{6}Department of Astronomy, Indiana University, 727 East Third Street, Bloomington, IN 47405, USA \quad
\textsuperscript{7}Instituto de Estudios Astrof\'isicos, Facultad de Ingenier\'ia y Ciencias, Universidad Diego Portales, Av.\ Ej\'ercito 441, Santiago, Chile \quad
\textsuperscript{8}Instituto de Astrof\'isica e Ci\^encias do Espa\c{c}o, Portugal \quad
\textsuperscript{9}Faculdade de Ci\^encias da Universidade de Lisboa, Portugal \quad
\textsuperscript{10}University of the Pacific, Stockton, CA 90340 USA \quad
\textsuperscript{11}Univ Lyon, Univ Lyon1, Ens de Lyon, CNRS, CRAL UMR5574, F-69230, Saint-Genis-Laval, France \quad
\textsuperscript{12}Armagh Observatory and Planetarium, College Hill, Armagh, BT61 9DG, N. Ireland, UK \quad
\textsuperscript{13}Centre for Astrophysics Research, Department of Physics, Astronomy and Mathematics, University of Hertfordshire, Hatfield AL10 9AB, UK \quad
\textsuperscript{14}Institute for Computational Cosmology, Department of Physics, Durham University, South Road, Durham DH1 3LE, UK \quad
\textsuperscript{15}University of Ljubljana, Faculty of Mathematics and Physics, Jadranska ulica 19, SI-1000 Ljubljana, Slovenia \quad
\textsuperscript{16}Department of Physics and Astronomy, University of California Davis, 1 Shields Avenue, Davis, CA 95616, USA \quad
\textsuperscript{17}Columbia University \quad
\textsuperscript{18}Kavli Institute for Cosmology, University of Cambridge, Madingley Road, Cambridge CB3 0HA, UK \quad
\textsuperscript{19}Institute of Astronomy, University of Cambridge, Madingley Road, Cambridge CB3 0HA, UK \quad
\textsuperscript{20}School of Earth and Space Exploration, Arizona State University, Tempe, AZ 85287, USA \quad
\textsuperscript{21}Cosmic Dawn Center (DAWN), Copenhagen, Denmark \quad
\textsuperscript{22}Niels Bohr Institute, University of Copenhagen, Jagtvej 128, DK-2200 Copenhagen, Denmark \quad
\textsuperscript{23}Scuola Normale Superiore, Piazza dei Cavalieri 7, I-56126 Pisa, Italy \quad
\textsuperscript{24}David A.\ Dunlap Department of Astronomy \& Astrophysics, University of Toronto, Toronto, ON M5S 3H4, Canada \quad
\textsuperscript{25}INAF -- Osservatorio Astronomico di Roma, Via Frascati 33, I-00078 Monte Porzio Catone, Italy \quad
\textsuperscript{26}Dipartimento di Fisica, Universit\`a di Roma Tor Vergata, Via della Ricerca Scientifica 1, I-00133 Roma, Italy \quad
\textsuperscript{27}INFN, Sezione di Roma I, Piazzale Aldo Moro 2, I-00185 Roma, Italy \quad
\textsuperscript{28}Institute for Astronomy, University of Edinburgh, Royal Observatory, Edinburgh EH9 3HJ, UK \quad
\textsuperscript{29}Centre for Astrophysics and Supercomputing, Swinburne University of Technology \quad
\textsuperscript{30}NSF National Optical-Infrared Astronomy Research Laboratory, 950 N. Cherry Ave., Tucson, AZ 85719, USA \quad
\textsuperscript{31}Department of Physics and Astronomy, University of Hawaii, Hilo, 200 W Kawili St, Hilo, HI 96720, USA \quad
\textsuperscript{32}Institut d'astrophysique de Paris, UMR 7095, CNRS, Sorbonne Universit\'e, 98 bis boulevard Arago, 75014 Paris, France \quad
\textsuperscript{33}IPAC, California Institute of Technology, 1200 E. California Blvd. Pasadena, CA 91125, USA \quad
\textsuperscript{34}Institute of Physics, Laboratory for Galaxy Evolution and Spectral Modelling, EPFL, Observatoire de Sauverny, Chemin Pegasi 51, CH-1290 Versoix, Switzerland \quad
\textsuperscript{35}Department of Astronomy, The University of Texas at Austin, Austin, TX 78712, USA \quad
\textsuperscript{36}Cosmic Frontier Center, The University of Texas at Austin, Austin, TX 78712, USA \quad
\textsuperscript{37}DTU Space, Technical University of Denmark, Elektrovej 328, DK-2800 Kgs.\ Lyngby, Denmark \quad
\textsuperscript{38}Institut f\"ur theoretische Astrophysik, Zentrum f\"ur Astronomie der Universit\"at Heidelberg, Albert-Ueberle-Str.\ 2, D-69120 Heidelberg, Germany \quad
\textsuperscript{39}Institute for Astronomy, University of Hawaii, 2680 Woodlawn Drive, Honolulu, HI 96822, USA \quad
\textsuperscript{40}U.S. National Science Foundation (NSF), Alexandria, VA 22314, USA \quad
\textsuperscript{41}Instituto de Astrof\'isica de Canarias, c/ V\'ia L\'actea s/n, E-38205 La Laguna, Tenerife, Spain \quad
\textsuperscript{42}Departamento de Astrof\'isica, Universidad de La Laguna, E-38205 La Laguna, Tenerife, Spain \quad
\textsuperscript{43}Department of Astrophysics, University of Vienna, T\"urkenschanzstr. 17, 1180 Vienna, Austria \quad
\textsuperscript{44}University of Toronto, Toronto, ON M5S 3H4, Canada \quad
\textsuperscript{45}Department of Physics and Astronomy, University of Maine, Orono, ME 04469, USA \quad
\textsuperscript{46}Max Planck Institute for Astrophysics \quad
\textsuperscript{47}Department of Astronomy, Yonsei University, 50 Yonsei-ro, Seodaemun-gu, Seoul 03722, Republic of Korea \quad
\textsuperscript{48}Space Telescope Science Institute, 3700 San Martin Drive, Baltimore, MD 21218, USA \quad
\textsuperscript{49}SISSA, Via Bonomea 265, 34136 Trieste, Italy \quad
\textsuperscript{50}Physics Department, College of Science, United Arab Emirates University, PO Box 15551, Al-Ain, UAE \quad
\textsuperscript{51}Department of Physics, University of California, Santa Barbara, CA 93106, USA \quad
\textsuperscript{52}INAF - Osservatorio Astrofisico di Arcetri \quad
\textsuperscript{53}Department of Physics and Astronomy, University of Catania, Via Santa Sofia 64, 95123 Catania, Italy \quad
\textsuperscript{54}The University of Osaka \quad
\textsuperscript{55}Flatiron Institute \quad
\textsuperscript{56}Department of Astronomy, Yale University, Kline Tower, 266 Whitney Avenue, New Haven, CT 06511, USA \quad
\textsuperscript{57}Department of Physics, Yale University, P.O. Box 208121, New Haven, CT 06520, USA \quad
\textsuperscript{58}Black Hole Initiative at Harvard University, 20 Garden Street, Cambridge, MA 02138, USA \quad
\textsuperscript{59}Department of Astronomy \& Oskar Klein Centre for Cosmoparticle Physics, AlbaNova, Stockholm University, SE-106 91 Stockholm, Sweden \quad
\textsuperscript{60}Centro de Astrobiolog\'{\i}a (CAB), CSIC-INTA, Ctra. de Ajalvir km 4, Torrej\'on de Ardoz, E-28850, Madrid, Spain \quad
\textsuperscript{61}Dunlap Institute for Astronomy \& Astrophysics, University of Toronto, Toronto, Ontario, Canada \quad
\textsuperscript{62}Konkoly Observatory \quad
\textsuperscript{63}Instituto Galego de F\'isica de Altas Enerx\'ias, Universidade de Santiago de Compostela, 15782 Santiago de Compostela, Galicia, Spain \quad
\textsuperscript{64}University of Bath, Department of Physics, Claverton Down, Bath, BA2 7AY, UK \quad
\textsuperscript{65}School of Physics and Astronomy, University of Southampton, Highfield, Southampton, SO17 1BJ, UK \quad
\textsuperscript{66}Kavli Institute for Cosmological Physics, University of Chicago, Chicago IL 60637, USA \quad
\textsuperscript{67}University of Oxford \quad
\textsuperscript{68}Department of Physics and Astronomy, University of Kentucky, 505 Rose Street, Lexington, KY 40506, USA \quad
\textsuperscript{69}University of California, Los Angeles, 475 Portola Plaza, Los Angeles, CA, 90095, USA \quad
\textsuperscript{70}CIERA and Department of Physics and Astronomy, Northwestern University \quad
\textsuperscript{71}Astronomy Centre, University of Sussex, Falmer, Brighton BN1 9QH, UK \quad
\textsuperscript{72}School of Astronomy and Space Science, Nanjing University, Nanjing, Jiangsu 210023, China \quad
\textsuperscript{73}Georgia Institute of Technology \quad
\textsuperscript{74}National Tsing Hua University - Taiwan \quad
\textsuperscript{75}Kavli Institute for Particle Astrophysics \& Cosmology, Stanford University, 452 Lomita Mall, Stanford, CA 94305, USA \quad
\textsuperscript{76}Department of Physics, Stanford University, 382 Via Pueblo Mall, Stanford, CA 94305, USA \quad
\textsuperscript{77}University of Illinois Urbana-Champaign}\par}

\clearpage
\bibliography{rss}
\bibliographystyle{aasjournalv7.1.bst}

\end{document}